\documentclass[10pt,twocolumn]{article}
\usepackage{times,mathptmx}

\usepackage{subfigure}
\usepackage{xspace}

\usepackage{color}			
\usepackage{verbatim}		
\usepackage{latexsym}		
\usepackage{mathrsfs}		
\usepackage{graphicx}		
\usepackage{qtree}			
\usepackage{listings}		
\usepackage{mdframed}		
\usepackage{float}			
\usepackage{scrextend}		
\usepackage{caption}		
\usepackage{hhline}			
\usepackage[hyphens]{url}
\usepackage{amsmath} 		
\usepackage{siunitx}

\begin{document}





\title{SIVSHM: Secure Inter-VM Shared Memory}

\author{
Shesha B. Sreenivasamurthy \\
\textit{Univ. of California, Santa Cruz} \\
\textit{shesha@ucsc.edu}
\and
Ethan L. Miller \\
\textit{Univ. of California, Santa Cruz} \\
\textit{elm@cs.ucsc.edu}
}
\date{}

\maketitle
\thispagestyle{empty}


\begin{abstract}
With wide spread acceptance of virtualization, virtual machines (VMs) find their presence in various applications such as Network Address Translation (NAT) servers, firewall servers and MapReduce applications. Typically, in these applications a data manager collects data from the external world and distributes it to multiple workers for further processing. Currently, data managers distribute data with workers either using inter-VM shared memory (IVSHMEM) or network communication. IVSHMEM provides better data distribution throughput sacrificing security as all untrusted workers have full access to the shared memory region and network communication provides better security at the cost of throughput. Secondly, IVSHMEM uses a central distributor to exchange \textit{eventfd} -- a file descriptor to an event queue of length one, which is used for inter-VM signaling. This central distributor becomes a bottleneck and increases boot time of VMs. Secure Inter-VM Shared Memory (SIVSHM) provided both security and better throughout by segmenting inter-VM shared memory, so that each worker has access to segment that belong only to it, thereby enabling security without sacrificing throughput. SIVSHM boots VMs in 30\% less time compared to IVSHMEM by eliminating central distributor from its architecture and enabling direct exchange of eventfds amongst VMs.
\end{abstract}

\section{Introduction}
\label{sec:intro}
In a cloud computing, a collection of VMs provide a service. In a multi-tenant environment, each tenant runs multiple such services. MapReduce services will have a manager (mapper) that farms incoming data to multiple workers (reducers), making routing decisions, and the workers run the computation using programs provided by untrusted third party computation providers and return the result back to the manager \cite{18}. In the rest of this paper, manager and mapper are used interchangeably and so are workers and reducers. The goal is to ensure that each reducer has access to only its portion of data thereby preventing any information leak amongst the reducers. Data sharing using traditional network provides that natural boundary whereby data of one reducer is inaccessible to another. If mapper and reducer VMs run on the same physical host, data can be shared by inter-VM shared memory \cite{11, 13}. Although this improves performance significantly, it is the vulnerable to information leakage among the VMs.

SIVSHM solves this problem by segmenting the shared memory and mapping only a segment to each reducer VM in the hypervisor. Thus, each reducer VM has access to only its segment. Any illegal access of memory by the guest kernel impacts only the adversary VM without affecting host, mapper or other reducer VMs.

Inter-VM interrupts are used between mapper and reducer VMs to signal data availability and task completion. This is accomplished by exchanging eventfds \cite{20, 21} during startup with the aid of an eventfd distributor \cite{11, 13}. One eventfd distributor per service is required for services using IVSHMEM. However, in a multi-tenant multi-service environment, eventfd distributor becomes a bottleneck during VM startup, thereby increasing boot time. Additionally, this extra software component needs to be managed by cloud service management software and is a management overhead.

SIVSHM makes inter-VM shared memory more conducive to the cloud environment by enabling direct exchange of eventfds between mapper and reducers thereby eliminating eventfd distributor from the architecture. This enables SIVSHM to boot a service with 32 VMs in 30\% less time compared to IVSHMEM \cite{11, 13}. SIVSHM makes the following two contributions:
\begin{enumerate}
	\item secure inter-VM shared memory architecture
	\item improved boot-time of services
\end{enumerate}

SIVSHM is built using the same underlying mechanism (virtual-PCI device) as that of IVSHMEM and therefore performs similarly during data transfer as shown in figure~\ref{times}. The advantages of SIVSHM over IVSHMEM is security, improved boot time and manageability. Data between mapper and reducers are predominantly transferred using high speed network interfaces. Therefore, throughput of SIVSHM is compared with VirtIO \cite{1} and to the best of our knowledge no prior work has made such a comparative study in MapReduce context. Compared to VirtIO, SIVSHM takes 40\% less time to process 32 GB of data for small configuration with 3  reducers and 60\% less time for large configuration with 31 reducers.

\section{Background}
\label{sec:background}
In general, MapReduce applications fall into Recognition, Mining and Synthesis (RMS) framework proposed by Intel \cite{10}. They find their presence in various applications such as database engines, virtual routers, virtual firewalls, load balancers, video stream processing and numerous other applications including neural prostheses \cite{9}. There can be applications where intermediate results produced on one machine are processed on another, either simultaneously or later in time \cite{12}. SIVSHM can be used by all applications that fall under this category.

There have been efforts to improve the performance of MapReduce type of applications. Phoenix \cite{2} is an implementation of MapReduce for shared-memory systems that includes a programming API and an efficient runtime system. However, the drawback of Phoenix is, both mapper and reducers run natively on physical system, which makes it less conducive to cloud environment. SIVSHM solves this problem by enabling mapper and reducers to run on different VMs and also opens opportunity to have multi-operating system environment where reducer applications can be running on different types of operating systems. Additionally, failure of mapper or a reducer VM will or affect other VMs in the system and the failed VM can be restarted without unlinking the shared memory thereby not losing any running VMs' work.

Similar to SIVSHM, IVSHMEM can be used to share memory between VMs. Both are implemented as virtual-PCI (vPCI) devices in QEMU (explained later) and map shared memory region to PCI device memory.  Data transfer performance of SIVSHM and IVSHMEM will be similar as they are built using the same underlying vPCI mechanism. However, the advantage of SIVSHM over IVSHMEM is security, improved boot time and manageability.

Non IVSHMEM/SIVSHM MapReduce services distribute data between mapper and reducers over the network using one of the two popular virtual network devices -- e1000 or VirtIO. VirtIO is a virtual network device that enables high speed data transfer between any two VMs. Simple \textit{iperf} \cite{8} test shows VirtIO has 10x bandwidth compared to virtual e1000 device (4.12 Gbps Vs 406 Mbps). As VirtIO is the predominant way to transfer data between VMs, data transfer throughput of SIVSHM is compared with VirtIO and IVSHMEM.

Several applications have previously been developed using the IVSHMEM infrastructure \cite {15, 16, 17}. Gordon has modified Phoenix MapReduce application to use IVSHMEM so that each Phoenix MapReduce thread runs inside a VM. This enables Phoenix to use shared memory and yet be conducive to cloud environment \cite {14}. However, it suffers from information leakage as explained earlier. 

Airavat \cite{18} runs on SELinux \cite{19} to provide security to MapReduce applications in cloud environment without shared memory. Data is exchanged over the network and therefore would perform similar to VirtIO. SIVSHM improves performance with the use of shared memory and provides security by memory segmentation.

Quick EMUlator (QEMU) \cite{3, 4} is a user space hardware emulator combined with Kernel Virtual Machine (KVM) \cite{6} provides complete virtualization environment in Linux. Hardware devices are emulated in QEMU while memory management and guest instruction execution is performed by KVM. QEMU triggers guest instruction execution using a blocking \textit{KVM\_RUN} ioctl to KVM. When guest I/O instruction such as reading and writing to a vPCI IO-registers are encountered by KVM, it returns from ioctl for QEMU to emulate those hardware operations. During the time when control is within QEMU, guest instruction execution has stopped. Similarly, when a user signal is sent to QEMU process, KVM returns from ioctl and after handling the signal, guest instruction execution is resumed by  QEMU by issuing \textit{KVM\_RUN} ioctl. We can notice that a read/write to vPCI IO-registers by the guest, results in a VM-exit and a context switch, which are expensive operations \cite{25, 26}.

Eventfd forms the backbone of both IVSHMEM and SIVSHM's interrupt architecture. The \textit{eventfd()} system-call creates a kernel object that can be used as an event wait/notify mechanism between user-space applications and kernel. The system-call returns a file descriptor (\textit{fd}) associated with the kernel object called \textit{eventfd} to the user application. The kernel object contains an unsigned 64-bit integer counter that is maintained by the kernel. In IVSHMEM and SIVSHM, eventfds are used by QEMU -- a user-space application and host-kernel. Eventfds are mapped to vPCI IO-registers by QEMU, which enables the hypervisor (host-kernel) to directly notify VM$_b$ while executing guest instructions of VM$_a$. These \textit{fd}s are exchanged between VMs for efficient inter-VM signaling during startup, by placing them in the control field of a unix-socket message. Placing \textit{fd}s in the control field of a message, is the standard linux way of \textit{fd} exchange, whereby the sender process (QEMU$_a$) is instructing the host-kernel to ensure that the receiving process (QEMU$_b$) receives a \textit{fd} unique in its name space. Eventfd exchange mechanism among VMs is improved in SIVSHM (explained in the next section), but the core signaling mechanism is very similar to IVSHMEM and we refer the reader to IVSHMEM \cite{13} for detailed explanation.

All applications developed using IVSHMEM can also be developed using SIVSHM without having any of the above mentioned drawbacks. In this work, a simple IO intensive MapReduce application is chosen to demonstrate its feasibility.

\section{Design}
\label{sec:design}
\begin{figure*}
    \begin{center}
        \includegraphics[width=0.75\textwidth, height=6.3cm]{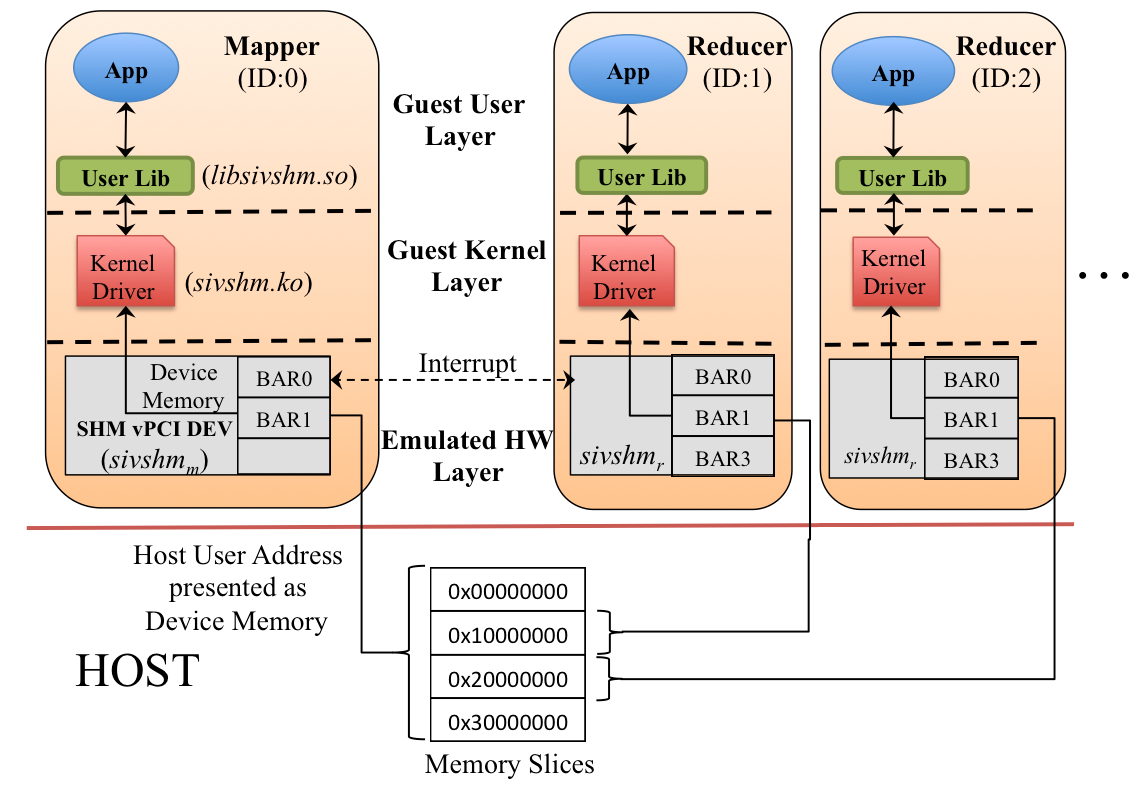}
        \captionsetup{justification=centering,font=footnotesize}
        \caption{SIVSHM Architecture \label{sivshm}}
    \end{center}
\end{figure*}

SIVSHM architecture contains a trusted mapper VM and a farm of untrusted reducer VMs. Guest user-space mapper and reducer applications run inside mapper and reducer VMs respectively. SIVSHM architecture is as shown in figure~\ref{sivshm}. Polling or interrupt mechanism is used between mapper and reducers to signal data availability and task completion. Mapper application stripes the data received from external world to respective memory slices of the reducers and signal the reducers of data availability. The reducers process the data by accessing their respective slice and inform the mapper when their task is completed. Mapper consolidates the data and presents the result. The main advantages of SIVSHM, security and improved boot performance are explained below.

\textbf{Security:} To share memory among VMs and for inter-VM signalling, we implemented a vPCI device called ``\textit{sivshm}" in QEMU. Mapper is always the first VM to be instantiated, which gets ID 0 and reducers are assigned non-zero IDs. Mapper's \textit{sivshm} device (\textit{sivshm$_m$}) creates a shared memory region and maps the entire region to its PCI device memory. In IVSHMEM, reducer too maps the entire shared memory region similar to mapper. In contrast to IVSHMEM, reducer's \textit{sivshm} device (\textit{sivshm$_r$}) in SIVSHM gets shared memory ID and the size of its memory slice from \textit{sivshm$_m$}. It offsets into the shared memory region using its own ID as the key and maps only its slice to the PCI device memory before booting the guest. This ensures that guest running inside reducer VM has access to only its slice, thereby providing superior security over IVSHMEM shared memory architecture.

\begin{figure}
    \begin{center}
        \includegraphics[width=0.5\textwidth]{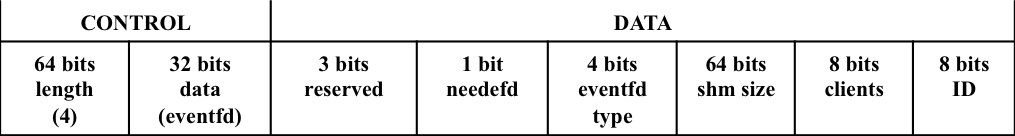}
        \captionsetup{justification=centering,font=footnotesize}
        \caption{Format of the message exchanged between \textit{sivshm$_m$} and \textit{sivshm$_r$} during startup\label{pdu}}
    \end{center}
\end{figure}

\textbf{Boot performance:} Inter-VM signaling between VMs is achieved by eventfds\cite{20, 21} in both SIVSHM and IVSHMEM. In IVSHMEM, a mapper and reducers exchange eventfd through an eventfd distributor (man in middle). Eventfd distributor becomes a bottleneck when a service containing large number of reducers is instantiated. SIVSHM removes this bottleneck by direct exchange of eventfd between mapper and reducer VMs. Mapper and reducer VMs exchange information such as, shared memory size, number of clients and client ID via a unix-socket messages. Eventfds are piggybacked on those messages by placing them in the message's control field. The format of the message exchanged is as shown in figure~\ref{pdu}. The exchanged eventfds are added to their respective poll list, waiting to be notified by the hypervisor of any events. This design eliminates eventfd distributor used by IVSHMEM from SIVSHM's architecture enabling better boot performance as shown in section~\ref{sec:results}.

\section{Implementation}
\label{sec:implementation}
System implementation is explained by taking a bottom-up approach. Physical PCI/PCI-E devices have an on board RAM called device memory. The address of the device memory is provided in PCI register called Base Address Register (\textit{BAR}). PCI standard  \cite{24} supports 6 BARs and SIVSHM uses two of them -- \textit{BAR0} for IO register space, \textit{BAR1} for mapping its own slice. A vPCI device in QEMU emulates device memory by allocating host memory. In \textit{sivshm}, the vPCI device memory is a shared memory object created using \textit{MAP\_SHARED} flag. The shared memory is \textit{LOCKED} to prevent it from being swapped out along with the VM by the host OS as it is not just used by one VM but shared by many. The address of the shared memory object is provided in the emulated BAR. The address in the BAR is perceived as physical address by the guest that is mapped to kernel virtual address by the guest driver.

A guest kernel driver (\textit{sivshm.ko}) to drive this new vPCI device was implemented, that claims this device and maps the vPCI device memory address to the guest kernel virtual address. The user applications can use this device to map the device memory to user address space, retrieve device information and to generate interrupts. To aid of user applications, we implemented a shared library (\textit{libsivshm.so}) that hides the driver interface details from the user applications and exposes a simple API to user applications.

\textit{sivshm} requires a shared memory ID, a unix socket path, VM ID, size and maximum number of reducers as its input. Size and number of reducers are only used by the mapper. These parameters are implemented as device specific variables, which are specified as command line arguments to QEMU. Mapper is instantiated first and is always assigned an ID 0. Specified shared memory ID is deleted (if it is present in the system) and recreated. Optionally, \textit{unlink=0} can be passed instructing \textit{sivshm} not to delete it. Reducers are assigned a non-zero ID. \textit{sivshm$_m$} communicates the size of the slice to \textit{sivshm$_r$} over the unix socket.

The shared memory region is sliced into equal sized segments: \textit{slice\_size = total\_shm\_size / Total VMs}. \textit{Total VMs} includes mapper as it gets a slice too, which is used as a message box during polling. The design does not preclude reducers from having different sized segments. However, for the  workload used in our experiments, it was apt to have equal sized segments. \textit{sivshm$_r$} uses its own ID ($r_j$) as key to calculate the start address ($s_i$) of device memory region: $s_i = shm\_start\_address + (r_j \times slice\_size)$. $s_i$ is mapped to  \textit{BAR1} of \textit{sivshm$_r$} that allows reducer $r_j$ to access only its slice. A predefined location in the reducers' slice is used as mailbox to exchange signals between mapper and reducers in polling mode. 

All guest instructions are executed by the hypervisor on guests' behalf. A write by the guest to a vPCI IO-register mapped to an eventfd's kernel object, triggers an event to be delivered to the QEMU process waiting on the corresponding user level \textit{fd}. If mapper's guest writes to its IO-register, the hypervisor directly notifies \textit{sivshm$_r$}, which sets the interrupt bit. An interrupt is delivered immediately to the reducer VM via eventfd that gets handled by the guest vPCI driver. MapReduce applications register a signal (SIGUSR1) to be delivered when an interrupt is handled by the vPCI kernel driver. Similarly, reducers can also send signals to the mapper. 

When the VM boots, the guest kernel driver (\textit{sivshm.ko}) claims the \textit{sivshm} device. It maps the device memory to guest-kernel address and creates a device in the system device tree. Mapper and reducer guest applications communicate with the guest-kernel driver using simple APIs of SIVSHM's shared library (\textit{libsivshm.so}). Mapper signals the reducers after distributing the data so that the reducers can start processing it. Similarly, after reducers have completed their job, they signal the mapper to perform result consolidation. A pre-defined memory location in the shared memory region is monitored in polling mode for signal exchange. In interrupt mode, mapper and reducer guest-applications request the guest-kernel to send inter-VM interrupts by using \textit{sivshm\_notify} API. These applications register a callback handler with the guest-kernel driver, which are called when an interrupt is fired.

\section{Experiments and Results}
\label{sec:results}
The system is implemented using Linux 3.2.0-23 kernel as both host and guest operating system, KVM hypervisor and QEMU version 2.3.0 device emulator. The setup was tested on a system with 128 GB RAM and four 10-core processors with 2 hyper-threads per physical core, seen as 80 processing units by the operating system. Each VM was allocated 6 GB RAM. SIVSHM was experimented with one mapper VM and \{1, 3, 7, 15 and 31\} reducer VMs. Each vCPU was bound to a physical processing unit. Reducer VMs were allocated 2 vCPUs, one dedicated to the reducers' guest application and another one for kernel activities. Mapper VM runs a multi-threaded application that distributes data to multiple reducers. Hence, more vCPUs were allocated to it compared to reducers. With 31 reducer VMs, we have 62 CPUs allocated to reducers, 2 for the host hypervisor and the rest 16 to the mapper VM. 

\begin{figure}
    \begin{center}
        \includegraphics[width=0.5\textwidth, height=6cm]{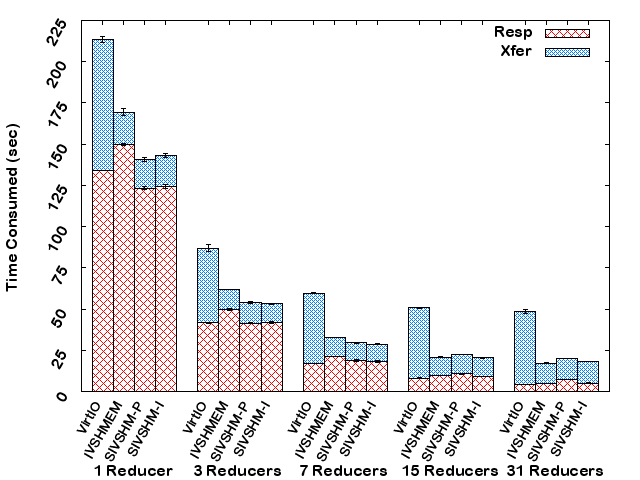}
        \captionsetup{justification=centering,font=footnotesize}
        \caption{Performance comparison of VirtIO, IVSHMEM and SIVSHM to process 32 GB of data with 1 GB shared memory.\\
        Xfer -- Time taken by all mapper threads to complete data transfer\\
        Resp -- Time the mapper waits to receive response from all reducers\label{times}\\
         `I' denotes Interrupt, `P' denotes Polling}
    \end{center}
\end{figure}

Entire set of experiments was repeated 10 times and the average of transfer and response times (explained later) is as shown in figure~\ref{times}. The error bars denote the standard deviation of the 10 values of each experiment. In each run, a total of 32 GB of random data was processed and performance of VirtIO, IVSHMEM and SIVSHM is compared. During VirtIO performance measurements, all VMs were networked using Linux bridge. Maximum possible bandwidth for VirtIO interfaces is achieved with this setup as no packets leave the compute node. 128 MB application buffer was allocated, as this was the maximum slice size that SIVSHM was experimented with.

We were interested in comparing IO performance. Therefore, a non-CPU intensive workload such as counting the number of occurrences of a specified character and storing the result in mapper's slice was chosen as our workload. Transfer time (\textit{Xfer}) in VirtIO is the time taken by all mapper threads to copy data from application buffer to kernel buffer. In SIVSHM and IVSHEMEM, it is the time to copy data from application buffer to reducer slices. Response time (\textit{Resp}) is the time the mapper waits to receive responses from all reducers.

In our experiments, 32 GB of data was processed with 1 GB shared memory region. Slice sized data is transferred by mapper in parallel to all reducers. Map-reduce performance increases as reducers are added because, amount of data processed by each reducer decreases. This in turn reduces the time taken to complete the work, especially response time, as shown in figure~\ref{times}. This effectively shows performance improvement due to parallelism.

The response time of SIVSHM and VirtIO are very similar as the workers in both the cases perform similar task -- counting number of occurrences of a character. However, transfer time of SIVSHM and IVSHMEM is significantly lower than VirtIO as both avoid a data copy from kernel buffer to VirtIO device queue in addition to TCP/IP stack overhead. Compared to VirtIO, SIVSHM takes (3/5)\textsuperscript{th} and (2/5)\textsuperscript{th} the time at low and high reducer counts in both polling and interrupt modes.

The response time of SIVSHM is marginally better than IVSHMEM at lower reducer count and the gap gradually decreases with higher reducer count. This is attributed to lower interrupt latency in SIVSHM --  amount of time between an interrupt is delivered to the device to the time the application handles that interrupt. SIVSHM has a kernel vPCI driver that signals the user process (SIGUSR1) when an interrupt is delivered to it. This is faster than IVSHMEM driver that is implemented using Linux UIO driver \cite{22} infrastructure, where an user-space process that is blocked on a read of a \textit{fd} has to be woken up. The interrupt latency of SIVSHM and IVSHMEM, on average, was measured to be 55 $\mu$sec and 120 $\mu$sec. Though the number of interrupts increase at higher reducer count, the latency gap decreases due to interrupt coalescing.

\begin{figure}
    \begin{center}
        \includegraphics[width=0.5\textwidth, height=6cm]{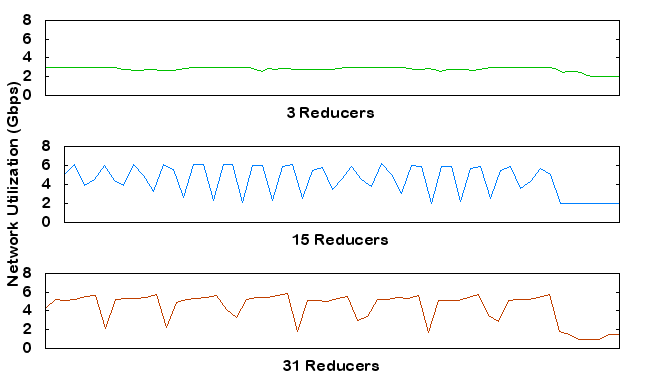}
        \captionsetup{justification=centering,font=footnotesize}
        \caption{VirtIO network performance (Mapper) \label{virtioperf}}
    \end{center}
\end{figure}

The transfer time remains approximately the same in all methods for different number of reducers. This is due to network saturation in VirtIO and memory bandwidth saturation in case of SIVSHM and IVSHMEM. In figure~\ref{virtioperf}, we can notice that, at higher reducer count the data transfer rate of VirtIO interface saturates at 6 Gbps. This saturation translates to constant transfer time when reducer count is greater than one. The performance drops multiple times as the mapper is waiting for all the reducers to process the data and report completion. For example -- a total of 32 GB of data is transferred from mapper to reducers and in case of 31 reducers, each reducer processes $\sim$1 GB of data and since the buffer size is 128 MB, we can notice in figure \ref{virtioperf} that the performance drops 8 times. 

\begin{figure}
    \begin{center}
        \includegraphics[width=0.35\textwidth, height=3.5cm]{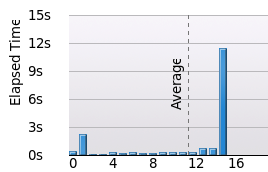}
        \captionsetup{justification=centering,font=footnotesize}
        \caption{Mapper CPU concurrency performance. x-axis denotes number of simultaneously utilized logical CPUs. Graph by Intel\textsuperscript{\textregistered}VTune\textsuperscript{TM}\cite{23}.\label{vtune}}
    \end{center}
\end{figure}

To substantiate our claim of memory bandwidth saturation, we show that limited concurrency and virtualization are not limiting factors, and only thing left -- memory bandwidth, should be the limiting factor. We ran the same mapper and reducer applications on the physical host and noticed that transfer time was less than 1\% better than running in a VM. This shows virtualization is not a limiting factor. Secondly, we measured concurrency of our mapper application using Intel's VTune Amplifier 2016 \cite{23}. The histogram produced by VTune with 15 mapper threads is as shown in figure~\ref{vtune}. The graph is denoting that 15 logical CPUs were simultaneously utilized for 12 seconds, which demonstrates high concurrency of our mapper application. 

\begin{figure}
    \begin{center}
        \includegraphics[width=0.5\textwidth, height=6cm]{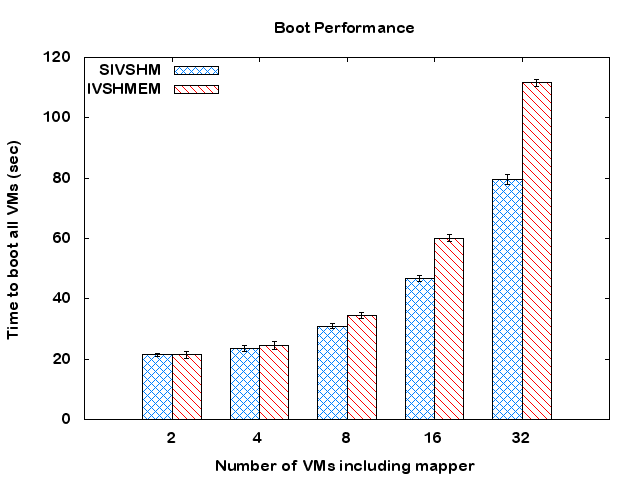}
        \captionsetup{justification=centering,font=footnotesize}
        \caption{Time to boot different number of VMs using SIVSHM and IVSHMEM. Shorter bar is better. \label{boot}}
    \end{center}
\end{figure}

We can infer from figure~\ref{times} that the improvement in transfer and response time of SIVSHM is not very significant when compared to IVSHMEM. This is because both are implemented as vPCI devices in QEMU, mapping shared memory region to PCI device memory. However, elimination of eventfd distributor in SIVSHM has eliminated eventfd exchange bottleneck during VM startup. This can be seen as improvement in boot-time in figure~\ref{boot}. At lower reducer count, no improvement is noticed as the number of eventfds exchanged is less. However, as the reducer count increases, we can notice that SIVSHM takes (7/10)\textsuperscript{th} the time to boot all the VMs when compared to IVSHMEM. This is significant, especially in a data centers where services are instantiated and torn down very frequently.

\begin{figure}
    \begin{center}
        \includegraphics[width=0.5\textwidth, height=6cm]{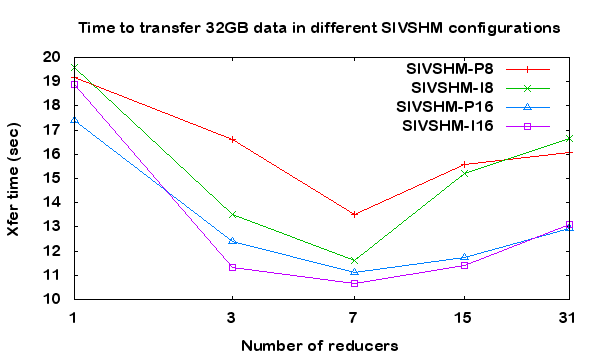}
        \captionsetup{justification=centering,font=footnotesize}
        \caption{SIVSHM performance with different number of vCPUs. `I' denotes Interrupt, `P' denotes Polling, 8 \& 16 denote number of vCPUs. Eg: SIVSHM-I8 --- SIVSHM in interrupt mode with 8 vCPUs. \label{times8vs16}}
    \end{center}
\end{figure}

Cloud pricing model is based on resource consumption. Therefore, we wanted to measure SIVSHM's performance with reduced vCPU count for the mapper. We reduced to 8 vCPUs, half of the original count. As expected, it can be inferred from figure~\ref{times8vs16} that SIVSHM performs better with increased vCPU count. With half the number of vCPUs, 1.25x increase in transfer time is noticed. Improvement is seen even when the number of reducers, and hence number of mapper threads, are less than number vCPUs. This is attributed to vCPU resource sharing mechanism in the hypervisor. However, the gap increases when the ratio of reducer count to vCPU count increases.

\begin{figure}
    \begin{center}
        \includegraphics[width=0.5\textwidth, height=5cm]{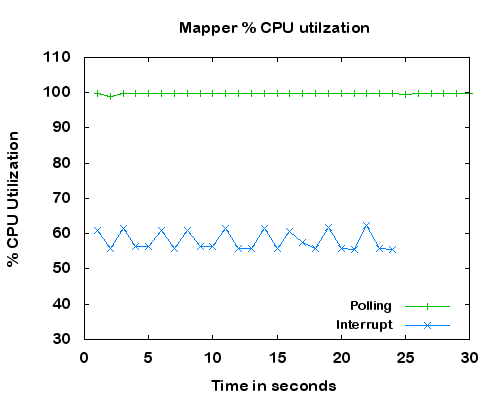}
        \captionsetup{justification=centering,font=footnotesize}
        \caption{CPU utilization between polling and interrupt \label{cpu}}
    \end{center}
\end{figure}

Another interesting observation is that SIVSHM shows similar performance in both polling and interrupt mode as seen in figure~\ref{times}. With similar performance, the motivation to implement more complex interrupt architecture is the improvement in CPU utilization -- 100\% in polling mode and hovers between 55-60\% in interrupt mode as shown in figure~\ref{cpu}. The remaining 40\% is available for other processes, which is a huge benefit of using interrupt driven architecture.
 
\section{Conclusion}
\label{sec:conclusion}
Shared memory is used to significantly boost the performance of regular applications from a long time. However, sharing memory among VMs is a recent thing. SIVSHM is a secure inter-VM shared memory architecture that can be used to boost performance of many cloud applications. SIVSHM takes 0.6x and 0.4x the amount of time compared to VirtIO for small and large number of reducers as VirtIO involves extra data copying and additional TCP/IP stack overhead.

The main restriction of SIVSHM is that the VMs should be running on the same compute node. In spite of this, applications can still be benefitted by SIVSHM as many of these applications run multiple copies for performance and to insulate customers from software bugs, both of which can be achieved by running VMs carrying these applications on the same compute node. However, we believe that we can overcome this restriction if we build an architecture where SIVSHM is used when VMs are co-located on the same compute node and utilize RDMA \cite{7} technology when VMs are on different compute nodes. This hybrid architecture can then be used in any data center or cloud environment to improve the performance of variety of applications without the restriction that VMs should be running on the same compute node.

\end{document}